\documentclass[%
 reprint,
superscriptaddress,
 amsmath,amssymb,
 aps,
]{revtex4-2}
\usepackage{xcolor}
\usepackage{graphicx}
\usepackage{dcolumn}
\usepackage{comment}
\usepackage{bm}
\usepackage{float} 
\usepackage{placeins}

\begin{document}


\title{Exact eigenvalues and experimental signatures of Heisenberg-Kitaev interactions in spin-$\frac{1}{2}$ quantum clusters}

\author{E. M. Wilson}
\affiliation{%
Department of Physics, University of North Florida, Jacksonville, FL 32224, USA}

\author{Jian-Xin Zhu}
\affiliation{%
Theoretical Division and Center for Integrated Nanotechnologies, Los Alamos National Laboratory, Los Alamos, NM 87545, USA}

\author{J. T. Haraldsen}
\affiliation{%
Department of Physics, University of North Florida, Jacksonville, FL 32224, USA}


\date{\today}

\begin{abstract}

We investigate the thermodynamics and energy eigenstates of a spin-1/2 coupled trimer, tetramer in a star configuration, and tetrahedron. Using a Heisenberg Hamiltonian with additional Kitaev interactions, we explore the thermodynamic signatures of the Kitaev interaction. Our results show that introducing a Kitaev interaction generates a second Schottky anomaly in the heat capacity for systems with a large $K/J$ ratio. The Kitaev term also introduces nonlinear eigenvalues with respect to a magnetic field, pushing the clusters toward a regime similar to the incomplete Paschen-Back effect and triggering first and second-order quantum phase transitions along with robust thermodynamic behavior. Through this approach, we provide exact analytical solutions that offer insights into Kitaev interactions, both in molecular magnets and in extended systems such as honeycomb or Kagome lattices. Furthermore, we provide insight into experimental measurements for detecting Kitaev interactions in clusters.

\end{abstract}

\maketitle


\section{Introduction}

\begin{figure*}
    \centering
    \includegraphics[width= 6 in]{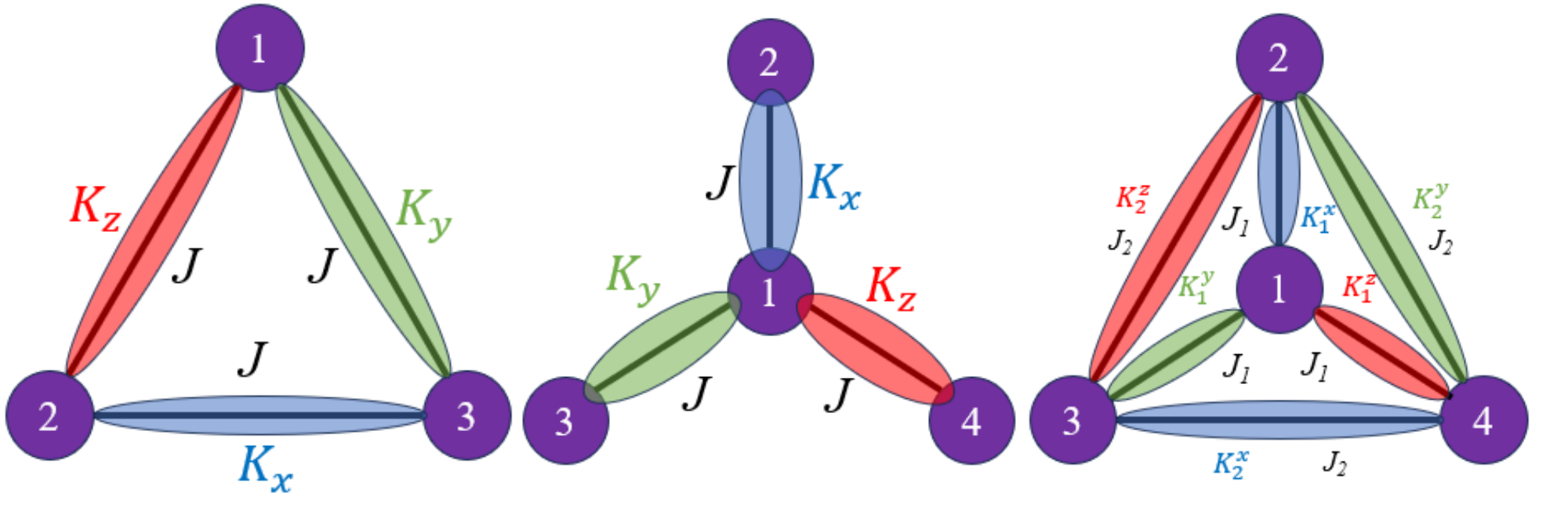}
    \caption{Diagram of a quantum spin coupled trimer, three-sided star tetramer, and tetrahedron cluster. Sites 1-4 represent spin 1/2 atoms that interact through isotropic J interaction (the black lines) and anisotropic interaction K (colored ellipses), which are directionally dependent.}
    \label{fig:clusterfigures}
\end{figure*}

Kitaev interactions have garnered significant attention in recent years due to their connection to possible quantum spin liquid states, Majorana fermions, and potential qubit applications \cite{taka19NatureReviews}. Experimental realizations of these states are hoped to provide a new and exciting leap in quantum technologies, such as quantum information conservation. However, while this has not yet been realized experimentally, many researchers are investigating the models in search of experimental signatures \cite{Rosl19InogChem}. 

Some suggested signatures appear in experimental techniques such as the Magneto-Optical Faraday Effect (MOFE) \cite{Nasu19PRR}. The production of a non-zero Faraday rotation angle, associated with finite transverse optical conductivity, has been claimed to signify a Kitaev-driven quantum spin liquid system. Moreover, a defect-induced orbital magnetization could be a signature of time-reversal symmetry breaking in Kitaev QSL \cite{Natori20PRL}. 
Furthermore, investigations into frustrated magnetic systems have shown that spin correlations vary with temperature. The reported deviations are asserted to be due to spinon pairing when there is an abrupt alteration in the temperature dependence of the nuclear spin-lattice relaxation, indicating the possibility of a phase transition to a QSL phase \cite{Hermann18ARCMP, KHATUA2PR}.

Over the years, many materials have been identified as potential Kitaev QSL \cite{yama17PRL,taka19NatureReviews, Rosl19InogChem,liu20PRL}. Among them are iridates \cite{taka19NatureReviews}, $\alpha$-RuCl$_3$ \cite{yama17PRL}, Na$_3$Co$_2$SbO$_6$ \cite{liu20PRL}, and many other materials. Each of these materials exhibits properties suggestive of Kitaev interactions, a unique form of magnetic interaction that arises from the spin-orbit coupling and crystal structure \cite{TREBST22PR}. In Kitaev QSL, the magnetic moments of the constituent atoms, typically transition metal ions, interact in a highly anisotropic manner dictated by the underlying lattice geometry. This anisotropic interaction leads to intriguing quantum phenomena, such as suppressing conventional magnetic ordering even at absolute zero temperature \cite{Li20PRR}. 

The thermodynamics of Kitaev materials provides valuable insights into their exotic properties. Traditional thermodynamic quantities, such as specific heat, entropy, and thermal conductivity, manifest unique signatures in the presence of quantum spin disorder \cite{KHATUA2PR}. Understanding these thermodynamic signatures is crucial for characterizing and identifying candidate materials that host Kitaev-driven phases \cite{Broholm20Science}. Moreover, exploring the thermodynamic behavior of these materials provides insight into the underlying quantum exchanges and their interactions, ultimately governing the emergence of phenomena in materials.

Spin clusters offer an effective experimental and theoretical platform for studying competing interactions in a controlled environment. Their finite size results in unique quantum effects that can be probed through spectroscopy or thermodynamic measurements \cite{Chilton2022ARMR}. Applying Heisenberg-Kitaev interactions to these clusters enables the exploration of the transition from localized magnetic behavior to collective phenomena, providing insight into the emergence of properties such as quantum spin liquids (QSL) and magnonic behavior in larger systems \cite{Srinivasan20PRR}. While Kitaev interactions are primarily investigated on the honeycomb lattice, the model also has the potential to significantly influence similar geometries found in spin clusters or molecular magnets.

Many experimental and theoretical studies have asserted the dominance of Kitaev (and off-diagonal $\Gamma$) interactions within specific configurations and candidate QSL materials \cite{Johnson15PRB, RAN17PRL, Lee20PRL,Churchill24PRL, Yang20PRL, Gohike24PRB}. Although other studies have shown that in many situations, the Heisenberg term plays a significantly more important role \cite{Li19PRB, samanta2024QM, Pandey22PRB, Liu22PRB, Wilson24PRB}. Our findings explore the interplay between Kitaev and Heisenberg interactions, highlighting their temperature- and field-dependent thermodynamic behavior. Through an exact diagonalization of the Hamiltonian model, we determine the exact energy eigenvalues, thermodynamic properties, and entropy, enabling us to identify distinct differences between the Heisenberg and Kitaev models for experimental verification. We find that this simple cluster model yields a similar double-hump in the heat capacity, which is essential due to the multiple order parameters within the model. Upon examining the energy levels, we find that the fractionalization of spin exchange produces nonlinear energy eigenstates, resulting in additional spin mixing. Furthermore, energy level repulsion is observed, arising from decoupling the bond directionality of the Kitaev terms in response to a magnetic field. Distinct Schottky anomalies associated with the Kitaev interaction are observed, and the evolution of the Hilbert space with respect to Kitaev interaction is analyzed across the trimer, tetramer, and tetrahedron spin configurations. The results suggest the possible existence of Kitaev-based molecular magnets and thermodynamic signatures of the Kitaev interaction and provide insights into quasiparticle-continuum-level repulsion in SOC systems.

\section{Heisenberg-Kitaev Model}

The Heisenberg model describes an isotropic interaction between neighboring spins, where the coupling energy depends on the spin alignment. This model captures the essence of magnetic interactions in many materials, particularly those with localized magnetic moments. However, it does not accurately describe all magnetic materials. It favors parallel alignment for ferromagnetic (FM) interactions and anti-parallel alignment for antiferromagnetic (AFM) interactions. Since the Hamiltonian is invariant under rotations in spin space, the total spin $S_{\text{tot}}$ becomes an appropriate quantum number to describe the system \cite{Schollwock2004}. 

The Kitaev term involves particles interacting through bond-directional, Ising-like connections that are very responsive to the bond orientation ($\alpha$=x,y,z) of each interaction, which leads to the fractionalization of the entangled spin-orbit states into fermionic degrees of freedom. These unique properties support the notion of a ground state quantum spin-liquid with either gapped or gapless features, effectively introducing the concept of Majorana fermions within a static $Z_2$ gauge field \cite{TREBST22PR}. The proposition establishes that the application of a magnetic field should couple to all spatial orientations of the spin, reconstructing the gapless QSL phase into a gapped QSL with non-Abelian topological order \cite{KITAEV06AoP,READ00PRB}.
The Heisenberg-Kitaev (H-K) model combines the effects of isotropic Heisenberg exchange and anisotropic
Kitaev interactions, offering a comprehensive approach
to investigate frustrated magnetic systems \cite{Pedro20PRB}. In order to identify key behaviors that reflect the individual contributions of the Heisenberg and Kitaev interactions, it is necessary to add a magnetic field term to perturb the spin states within a given arrangement \cite{Janssen16PRL}. The effective Hamiltonian for this model can be written as

\begin{equation}
H = \sum_{i \neq j}J_{ij}\vec{\sigma_i}\cdot\vec{\sigma_j} + \sum_{\substack{i \neq j \\ \alpha = x,y,z}}{K^{\alpha}_{ij}\sigma^{\alpha}_i}\cdot{\sigma^{\alpha}_j +  E_B\sum_{i \neq j} \sigma_{i}^{z}},
\label{HKwithFieldEquation}
\end{equation}

where $J_{ij}$ is the Heisenberg exchange energy, which promotes alignment or anti-alignment of neighboring spins depending on its sign, $K^{\alpha}_{ij}$ characterizes the anisotropic Kitaev coupling, selectively affecting individual spin components along the $\alpha = x, y, z$ directions, $\vec{\sigma_i}$ is the spin vector, $\vec{\sigma}^{\alpha}_j$ are the individual components of the spin vector, and $\alpha$ can be $x$, $y$, and $z$. The $E_B$ term is the magnetic field energy in the $\sigma_{i}^{z}$ spin direction.

The spin-\(\frac{1}{2}\) system is the fundamental quantum unit of angular momentum, crucial in modeling magnetism. In the Heisenberg Hamiltonian, it represents localized magnetic moments interacting via exchange coupling, leading to quantum fluctuations absent in classical spins. Its non-commutative algebra allows for superposition and entanglement, influencing ground states and excitations in spin chains and lattices \cite{Auerbach1994}. This behavior underlies phenomena like quantum phase transitions, spin liquids, and frustration in low-dimensional magnetic systems \cite{Broholm20Science}.

The competition between these interactions can lead to a range of magnetic ground states, spanning both conventional ordered phases, such as FM and AFM states, and more exotic disordered phases like QSL 
\cite{leumer2020JoCMP}. The directional dependence of the Kitaev coupling introduces frustration, particularly in lattice geometries such as honeycomb or triangular lattices, where the interactions cannot all be satisfied at once. This frustration results in complex quantum behavior, playing a crucial role in generating highly entangled quantum states without long-range magnetic order, such as in molecular magnets \cite{Gohike17PRL}.  

In particular, subtle nonlinearity in energy levels is attributed to Kitaev contributions arising from the coupling $J-K$ and the level repulsion of the coupling $K-E_B$, as described by the von Neumann-Wigner theorem \cite{von1929PZ}. Introducing anisotropy through a second parameter in the off-diagonal term induces the nonlinear response. When this coupling is subjected to a magnetic-field perturbation, level repulsion arises, preventing energy levels from crossing. The eigenstate structure applies to matrices of arbitrary size that can be decomposed into smaller $2^n \times 2^n$ blocks, such as $8 \times 8$ matrices and is true for all three cases \cite{von1929PZ,Bernier18PRA,Plumb2016NatPhy}.

The Heisenberg part of the energy is isotropic, and the use of symmetry can be exactly solved using the total spin $(S_{tot}$ of the system \cite{Harald16PRB}. The Kitaev model does not conserve total spin. However, the fractionalization of spin exchange allows for exact solutions when diagonalized \cite{Barghathi20PRR}. With the energy eigenstates $E_i$, the partition function $Z$, the heat capacity $C$, and the entropy $S_{ent}$ of the system can be calculated  \cite{Haraldsen05PRB}.

The heat capacity is easily obtained from the partition function, which is given by

\begin{equation}
    C = k_B \beta^2 \frac{d^2}{d\beta^2} \ln(Z),
\end{equation}

\noindent where $k_B$ is Boltzmann's constant and $\beta$ = $1/k_B T$. In any quantum system, heat capacity is essential for understanding spin interactions, quantum excitations, and phase transitions governing a given material. Observations are scrutinized by checking that the respective calculations accurately yield the correct entropy for a general spin system, which can then be compared to the expected zero-temperature entropy of the system by

\begin{equation}
S_{ent} = \int_0^\infty \frac{C}{\beta}d\beta = k_B \ln\left( \frac{N}{N_0} \right),
\end{equation}

\noindent where $N$ is the dimensionality of the full Hilbert space and $N_0$ is the degeneracy of the ground state \cite{Squire78TNS}. In the presence of small temperature variations, peaks in entropy correspond to increases in heat capacity. Determining how the ground state shifts with changes in system parameters allows for a connection between heat capacity peaks, eigenvalue degeneracies, and quantum phase transitions.

\section{Coupled Trimer}

\begin{figure*}
    \centering
    \includegraphics[width= 6 in]{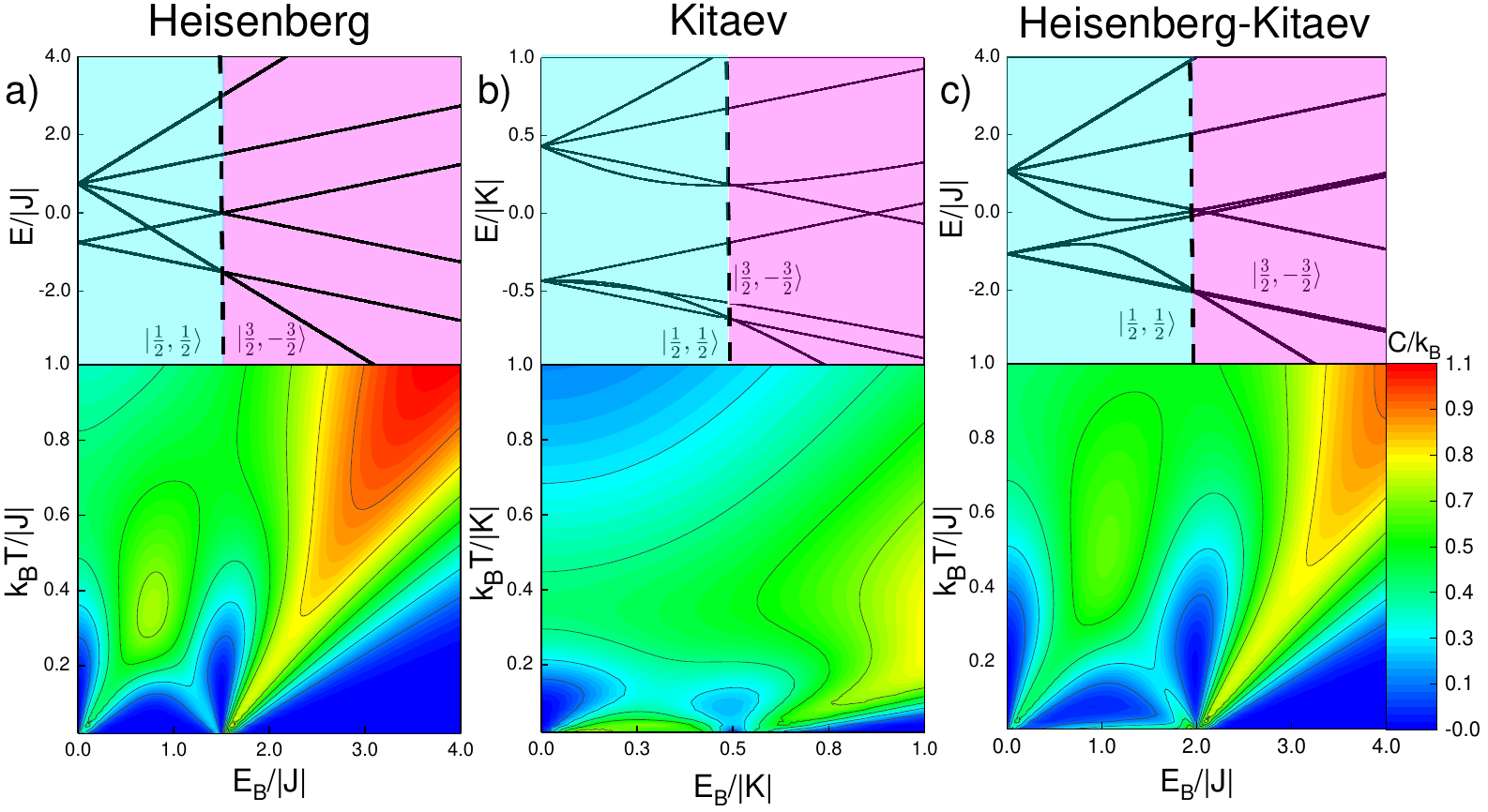}
    \caption{The exact energy eigenvalues with respect to the magnetic field and their corresponding temperature-dependent heat capacity for the coupled trimer are shown. A first-order phase transition from $S_{tot}$ = 1/2 to $S_{tot}$ = 3/2 is observed in all three cases. The Heisenberg interaction produces two distinct peaks in the heat capacity a). The Kitaev interaction results in less pronounced regions of spin mixing within the heat capacity b) that compete with the Heisenberg interaction in c). 
  }
    \label{fig:CoupledTrimerModel}
\end{figure*}

The unperturbed Heisenberg-Kitaev coupled trimer is shown in Fig.\ref{fig:clusterfigures}, and our effective Hamiltonian can be reduced to

\begin{equation}
\begin{array}{c}
H = J\Big(\vec{\sigma_1} \cdot \vec{\sigma_2} + \vec{\sigma_1} \cdot \vec{\sigma_3} + \vec{\sigma_2} \cdot \vec{\sigma_3} \Big) \\  
+K\Big( \sigma^{x}_2 \cdot \sigma^{x}_3 + \sigma^{y}_1 \cdot \sigma^{y}_3 + \sigma^{z}_1 \cdot \sigma^{z}_2  \Big),
\end{array}
\end{equation}

\noindent within the spin-1/2 basis, the coupled trimer has (2$S$+1)$^3$ = 8 magnetic states with the following spin decomposition from individual spin states to coupled spin states

\begin{displaymath}
    \frac{1}{2} \otimes \frac{1}{2} \otimes \frac{1}{2} = \frac{3}{2} \oplus \frac{1}{2}^2.
\end{displaymath}

\noindent Here, the superscripts indicate the number of spin states, where each state has a $2S + 1$ degeneracy. The perturbed Heisenberg-Kitaev spin matrix for the spin-$\frac{1}{2}$ coupled trimer is an 8x8 matrix produced as

\begin{widetext}
\begin{equation}
\resizebox{\columnwidth}{!}{$
H = \left(
\begin{array}{cccccccc}
\frac{3 J}{4} + \frac{K}{4} + \frac{3 \mathit{E_B}}{2} & 0 & 0 & \frac{K}{4} & 0 & -\frac{K}{4} & 0 & 0 \\
0 & -\frac{J}{4} + \frac{K}{4} + \frac{\mathit{E_B}}{2} & \frac{J}{2} + \frac{K}{4} & 0 & \frac{J}{2} + \frac{K}{4} & 0 & 0 & 0 \\
0 & \frac{J}{2} + \frac{K}{4} & -\frac{J}{4} - \frac{K}{4} + \frac{\mathit{E_B}}{2} & 0 & \frac{J}{2} & 0 & 0 & -\frac{K}{4} \\
\frac{K}{4} & 0 & 0 & -\frac{J}{4} - \frac{K}{4} - \frac{\mathit{E_B}}{2} & 0 & \frac{J}{2} & \frac{J}{2} + \frac{K}{4} & 0 \\
0 & \frac{J}{2} + \frac{K}{4} & \frac{J}{2} & 0 & -\frac{J}{4} - \frac{K}{4} + \frac{\mathit{E_B}}{2} & 0 & 0 & \frac{K}{4} \\
-\frac{K}{4} & 0 & 0 & \frac{J}{2} & 0 & -\frac{J}{4} - \frac{K}{4} - \frac{\mathit{E_B}}{2} & \frac{J}{2} + \frac{K}{4} & 0 \\
0 & 0 & 0 & \frac{J}{2} + \frac{K}{4} & 0 & \frac{J}{2} + \frac{K}{4} & -\frac{J}{4} + \frac{K}{4} - \frac{\mathit{E_B}}{2} & 0 \\
0 & 0 & -\frac{K}{4} & 0 & \frac{K}{4} & 0 & 0 & \frac{3 J}{4} + \frac{K}{4} - \frac{3 \mathit{E_B}}{2}
\end{array}
\right)
$}
\end{equation}
\end{widetext}

\begin{table}
\caption{Spin states $|S_{tot},S_z\rangle$ and their corresponding eigenvalues for the spin-$\frac{1}{2}$ trimer. Due to spin mixing with magnetic field, the $|\frac{3}{2},-\frac{3}{2}\rangle$ and one of the $|\frac{1}{2},\frac{1}{2}\rangle$ states flip.}
\centering
\begin{tabular}{|c|c|}
\hline
$|S_{tot},S_z\rangle$ & Eigenvalues \\ \hline

 $|\frac{3}{2},\frac{3}{2}\rangle$&
$ \frac{\mathit{E_B}}{2} + \frac{\sqrt{16 \mathit{E_B}^{2} + 24 J \mathit{E_B} + 8 K \mathit{E_B} + 9 J^{2} + 6 J K + 3 K^{2}}}{4} $  \\ \hline

$|\frac{3}{2},\frac{1}{2}\rangle$ & 
$ \frac{\mathit{E_B}}{2} + \frac{\sqrt{9 J^{2} + 6 J K + 3 K^{2}}}{4} $ \\ \hline

$|\frac{3}{2},-\frac{1}{2}\rangle$ &
$ -\frac{\mathit{E_B}}{2} + \frac{\sqrt{9 J^{2} + 6 J K + 3 K^{2}}}{4} $  \\ \hline

$|\frac{3}{2},-\frac{3}{2}\rangle$ ($|\frac{1}{2},\frac{1}{2}\rangle$)&
$ -\frac{\mathit{E_B}}{2} + \frac{\sqrt{16 \mathit{E_B}^{2} - 24 J \mathit{E_B} - 8 K \mathit{E_B} + 9 J^{2} + 6 J K + 3 K^{2}}}{4} $  \\ \hline

$|\frac{1}{2},\frac{1}{2}\rangle$ ($|\frac{3}{2},-\frac{3}{2}\rangle$) &
$ -\frac{\mathit{E_B}}{2} - \frac{\sqrt{16 \mathit{E_B}^{2} - 24 J \mathit{E_B} - 8 K \mathit{E_B} + 9 J^{2} + 6 J K + 3 K^{2}}}{4} $  \\ \hline

$|\frac{1}{2},-\frac{1}{2}\rangle$&
$ \frac{\mathit{E_B}}{2} - \frac{\sqrt{16 \mathit{E_B}^{2} + 24 J \mathit{E_B} + 8 K \mathit{E_B} + 9 J^{2} + 6 J K + 3 K^{2}}}{4} $  \\ \hline

$|\frac{1}{2},\frac{1}{2}\rangle$ & 
$ \frac{\mathit{E_B}}{2} - \frac{\sqrt{9 J^{2} + 6 J K + 3 K^{2}}}{4} $ \\ \hline

$|\frac{1}{2},-\frac{1}{2}\rangle$ &
$ -\frac{\mathit{E_B}}{2} - \frac{\sqrt{9 J^{2} + 6 J K + 3 K^{2}}}{4} $  \\ \hline

\end{tabular}%
\label{tab:trimer_eigenvalues}
\end{table}

We start by analyzing only the isotropic Heisenberg interactions that dominate the description of spin interactions in the majority of known materials. Because the magnetic interactions of the Heisenberg-coupled trimer have been thoroughly studied in Refs.\cite{brumfield2019thermodynamics, Haraldsen05PRB, Haraldsen09PRB, downing2021Proceedings, mentrup00PA}, we set the stage with this model to establish a foundation for comparing interactions.

As the magnetic field strength $E_B$ increases, spin mixing begins at $\frac{E_B}{|J|} = 0.75$, where the $S_{tot} = \frac{1}{2}$ states mix with the $S_{tot} = \frac{3}{2}$ states. At $\frac{E_B}{|J|} = 1.5$, we observe further mixing of $S_{tot} = \frac{1}{2}$ states, and more importantly, the ground state undergoes a first order phase transition from $S_{tot} = \frac{1}{2}$ to $S_{tot} = \frac{3}{2}$. In Fig. \ref{fig:CoupledTrimerModel} a), comparison with the heat capacity shows that the broad peak observed between $\frac{E_B}{|J|} = 0$ and $\frac{E_B}{|J|} = 1.5$ is associated with the superposition of spin states. The heat capacity increases until the onset of $S_{tot} = \frac{1}{2}$ state mixing, after which it begins to decline as the system approaches a quantum phase transition from $S_{tot} = \frac{3}{2}$  to $S_{tot} = \frac{1}{2}$ with the appropriate entropy of $S_{ent} = k_B \ln(2)$. After $\frac{E_B}{|J|} = 1.5$, the behavior of the heat capacity becomes smooth with respect to the field and temperature, which establishes our expectation of how the H-K model behaves when only the isotropic interactions are taken into account. 

Proceeding to a purely Kitaev trimer, we investigate the thermodynamics of a spin cluster dominated by anisotropic Ising-like interactions, which vary with the spatial orientation ($\alpha = x, y, z$) of each bond. As shown in Fig. \ref{fig:CoupledTrimerModel} (b), a key difference from the Heisenberg case is immediately apparent, as the Kitaev interactions result in nonlinear energy eigenvalues, and the degeneracy becomes fully lifted. The field required to induce spin mixing is much less than in the Heisenberg case and occurs from $\frac{E_B}{|K|}$ = 0 to $\frac{E_B}{|K|}$ = 1.0.

When the magnetic field increases in strength, the spin begins to decouple from the rotational momentum in one $S_{tot} = \frac{3}{2}$ state and two $S_{tot} = 0$  states. The two components ($ m_L $ and $ m_S $) start to precess around the magnetic field $E_B$ separately \cite{Ramos06TAJ,Plumb2016NatPhy}. The field strength surpasses the fine-structure separation between K-levels around $\frac{E_B}{|K|}$ = 0.1, the linear relationship no longer holds, and the eigenvalues enter a regime similar to the nonlinear incomplete Paschen-Back effect \cite{Stift08RAS,Modugno17PRA}. As the field goes beyond $\frac{E_B}{|K|}$ = 1.0, all nonlinear effects cease. 

The nonlinearity characteristics of the Kitaev eigenvalues are due to spin mixing. Most notably, the $S_{tot} = 0$ curves down with the field and mixes with $S_{tot} = \frac{1}{2}$  states before going linear as $S_{tot} = \frac{3}{2}$. At this point $\frac{E_B}{|K}$ = 0.5, the ground state transitions from a $S_{tot} = \frac{1}{2}$ to $S_{tot} = \frac{3}{2}$ state. When comparing the heat capacity, it is noticeable that there is no broad peak, as there is no singular crossover but rather a range of values within crossover proximity. The nonlinearity of the energy levels has a significant impact on the low-temperature dynamics of the heat capacity. Additionally, the less uniform shoulder that emerges at $\frac{E_B}{|K|} > 0.75$ corresponds to the degeneracy of the $S_{tot} = \frac{1}{2}$ states. The peak continues increasing with respect to the magnetic field as some eigenvalues approach the same energy levels. Having established the analytical thermodynamic behavior for both Heisenberg and Kitaev interactions separately, we set $K = J$ to symmetrically incorporate both interactions, forming the complete Heisenberg-Kitaev trimer. The energy eigenstates are provided in Table (\ref{tab:trimer_eigenvalues}).

\begin{figure*}
    \centering
    \includegraphics[width= 6 in]{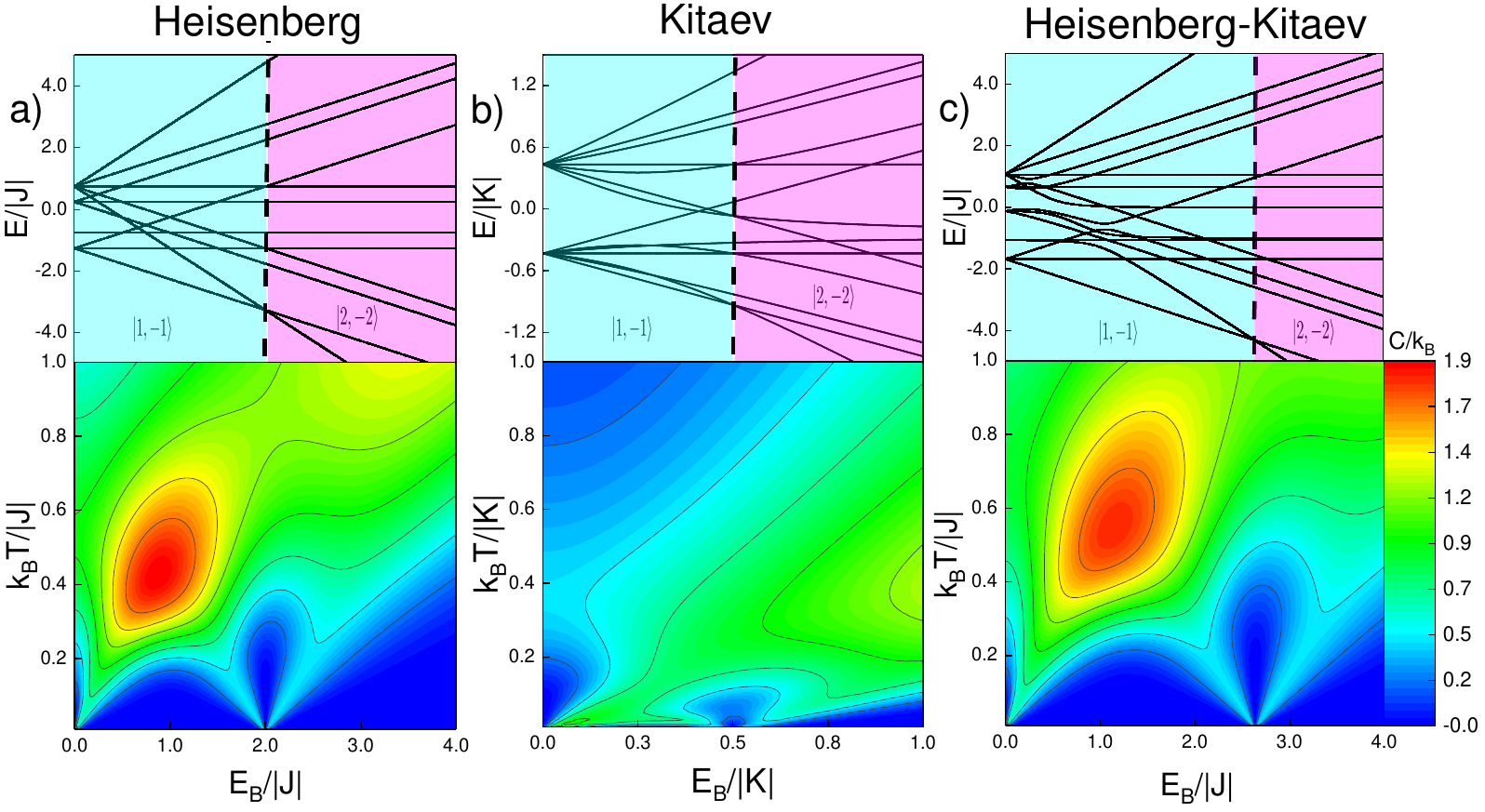}
    \caption{The exact energy eigenvalues with respect to the magnetic field and their corresponding temperature-dependent heat capacity for the three-sided star tetramer are shown. A first-order phase transition from $S_{tot}$ = 1/2 to $S_{tot}$ = 3/2 is observed in all three cases. The Heisenberg interaction produces a distinct peak in the heat capacity with a shoulder forming at high temperatures a). The Kitaev interaction results in pronounced regions of spin mixing with a shoulder appearing at higher fields within the heat capacity b) that enhances the effects of the Heisenberg interaction in c).}
    \label{fig:tetramerinteractions}
\end{figure*}

Looking back at Fig. \ref{fig:CoupledTrimerModel}, we observe that the overall shape of the energy levels is similar to the Heisenberg case. In the eigenstates shown in Table \ref{tab:trimer_eigenvalues}, the magnetic field $E_B$ does not appear in the coupling terms for the linear eigenstates, and the coupling between $J$ and $K$ does cause non-linear behavior but is drowned out by the magnetic field. In the nonlinear eigenstates, coupling terms involving $K$ and $E_B$, as well as $J$ and $E_B$ emerge, showing why the linearity returns at higher $E_B$. Moreover, the coupling between $K$ and $E_B$ is responsible for the energy level repulsion observed. The level repulsion introduced by the Kitaev interactions, coupled with the magnetic field, breaks the symmetry crossover point that would otherwise be present around $\frac{E_B}{|J|} = 0.75$. The nonlinear term curving downward with respect to the field is responsible for the phase transition from $S_{tot} = \frac{1}{2}$ to $S_{tot} = \frac{3}{2}$. The combination of both isotropic interactions and bond-dependent interactions increases the amount of field necessary for the quantum phase transition to occur. When compared to the heat capacity, we observe that the absence of the crossover causes the Heisenberg peak to broaden across higher temperatures. Although the isotropic behavior appears dominant, the orientation-dependent exchanges influence the finer details. This effect is most noticeable prior to the point of spin reconfiguration, where small additional peaks start to emerge.

\section{Three Sided Star Tetramer}

For the three-sided star model (shown in Fig. \ref{fig:clusterfigures}), the Hamiltonian can be reduced to

\begin{equation}
\begin{array}{c}
H = J\Big(\vec{\sigma_1}\cdot\vec{\sigma_2} + \vec{\sigma_1}\cdot\vec{\sigma_3} + \vec{\sigma_1}\cdot\vec{\sigma_4}\Big) \\  \\
+ K( \sigma^{x}_1\cdot\sigma^{x}_2 + \sigma^{y}_1\cdot\sigma^{y}_3 + \sigma^{z}_1\cdot\sigma^{z}_4.)
\end{array}
\end{equation}

Within the spin-1/2 basis, the tetramer has (2$S$+1)$^4$ = 16 magnetic states with following spin decomposition

\begin{displaymath}
    \frac{1}{2} \otimes \frac{1}{2} \otimes \frac{1}{2} \otimes \frac{1}{2} = 2 \oplus 1^3 \oplus 0^2.
\end{displaymath}

\noindent In the three-sided star structure, there is a considerable amount of spin mixing between states because of the shared spin in the center. Therefore, not all energy eigenstates are expected to be linear within the H-K model. The Heisenberg-Kitaev spin matrix for this configuration is a 16x16 matrix shown in Appendix A in equation \ref{eq:wideeq}. The eigenvalues associated with $S_{tot} = 2$ are the same as $S_{tot} = \frac{3}{2}$ and $S_{tot} = \frac{1}{2}$ in the trimer.

\begin{table}
\centering
\caption{The eigenstates $|S_{tot},S_z\rangle$ and eigenvalues for the three-sided star tetramer with no Kitaev interaction or no magnetic field.}
\begin{tabular}{|c|c|c|}
\hline
$|S_{tot},S_z\rangle$ & Eigenvalues & Eigenvalues\\
 & $K$=0 & $E_B$ = 0\\ \hline
$|2,S_z\rangle$  &  $\frac{3J}{4}+S_zE_B$ & $\frac{J}{2} + \frac{1}{4}\sqrt{J^2 + 2JK + 3K^2}$ \\

  &   & $\frac{1}{4}\sqrt{9J^2 + 6JK + 3K^2}$ \\ \hline

$|1,S_z\rangle$  & $\frac{J}{4}+S_zE_B$ & $-\frac{J}{2} + \frac{1}{4}\sqrt{9J^2 + 10JK + 3K^2}$ \\ \hline

$|1,S_z\rangle$  & $\frac{J}{4}+S_zE_B$ & $\frac{J}{2} - \frac{1}{4}\sqrt{J^2 + 2JK + 3K^2}$ \\ \hline

$|1,S_z\rangle$  & $-\frac{5J}{4}+S_zE_B$ & $-\frac{J}{2} - \frac{1}{4}\sqrt{9J^2 + 10JK + 3K^2}$ \\ \hline

$|0,0\rangle$  & $-\frac{3J}{4}$ & $-\frac{1}{4}\sqrt{9J^2 + 6JK + 3K^2}$ \\ \hline

$|0,0\rangle$  & $-\frac{3J}{4}$ & $-\frac{1}{4}\sqrt{9J^2 + 6JK + 3K^2}$\\ \hline

\end{tabular}%
\label{tab:star_tetramer_eigenvalues}
\end{table}

Following the same approach, the eigenvalues for the three-sided star tetramer are defined in Table \ref{tab:star_tetramer_eigenvalues}. The interactions are purely isotropic Heisenberg interactions, with no anisotropic exchange, while a magnetic field is applied. The spin mixing, induced by the central spin within the tetramer, lifts the degeneracy relative to the magnetic field, resulting in Zeeman splitting among 12 distinct energy levels. This spin mixing persists up to $\frac{E_B}{|J|} = 2.0$, where a transition from $S_{tot} = 1$ to $S_{tot} = 2$ occurs at the ground state level with a correct entropy of $S_{ent} = k_B \ln(2)$. This peak occurs within the range where the majority of energy level crossovers take place, corresponding to higher-order transitions. The isotropic interaction's contribution to the spin mixing generates a distinct conical peak in the heat capacity, emphasizing their influence. Compared to the trimer case, the ground-state transition spans a similar temperature range but requires a higher magnetic field strength. Moreover, the effect on the heat capacity is visually more symmetrical than the transition observed in the previous case. After the formal ground-state transition, no apparent Schottky anomaly is present; instead, the heat capacity contribution decreases steadily to higher temperatures after $\frac{E_B}{|J|}$ = 2.0.

We observe that the Kitaev three-sided star tetramer exhibits similar qualitative behavior to the Kitaev trimer in Fig. \ref{fig:tetramerinteractions} b). With the application of a magnetic field, nonlinear terms emerge immediately, placing the system within the incomplete Paschen-Back regime up to the ground-state transition at $\frac{E_B}{|J|}$ = 0.5. Beyond this range, the energy level splitting transitions to a linear pattern across the 14 observable levels, aligning with the expected behavior from the Zeeman effect. Repulsion between two positive and two negative eigenstates arises from the coupling between the bond-directional components and the magnetic field, driving a change in the ground state. Before the ground-state transition from $S_{tot} = 1$ to $S_{tot} = 2$, the contributions from the curved energy levels become evident at low temperatures. The higher-order partial degeneracies induce asymmetric crossovers, significantly lowering the field strength required to achieve substantial spin mixing, thereby removing the higher temperature peak observed in the trimer case and shifting its contributions to higher field values. Including additional nonlinear spin states enhances spin mixing within the energy levels, leading to exotic behavior in the heat capacity at lower temperatures and promoting a more well-defined phase transition.

With the characteristics of the Heisenberg and Kitaev interactions established, we examine how their behaviors intertwine. Unlike the trimer case, the coupling between the $J$ and $K$ terms does cause observable nonlinearity; however, it is similarly obscured by the coupling between $K$ and $E_B$. In this situation, the number of nonlinear terms is significantly higher than that of the pure Kitaev tetramer, and the energy level of the anti-crossing is much more pronounced. The competition among interactions further removes degeneracy from the system, revealing all 16 magnetic states, with half displaying linear behavior and the other half showing level repulsion, some on multiple occasions. As energy eigenstates undergo level repulsion, crossovers observed in the Heisenberg and Kitaev scenarios are forbidden, giving rise to increased and more complex spin mixing. Interestingly, the effect this has on the heat capacity is that the overall behavior observed in the Heisenberg context is spread out through high temperatures, and the field required to induce a quantum phase transition is increased to $\frac{E_B}{|J|}$ = 2.60, and there is no longer a peak at $\frac{E_B}{|J|}$ = 3.5. The overall heat capacity is more significant for the Kitaev and Heisenberg exchanges individually. Still, when together, the nonlinearity of the eigenstates reduces the maximum heat capacity. 

\begin{table}
\caption{Spin states $|S_{tot},S_z\rangle$ and their corresponding eigenvalues for the spin-$\frac{1}{2}$ symmetric tetrahedron. Due to spin mixing and the magnetic field, the $|2,-2\rangle$ and one of the $|0,0\rangle$ states flip. Here, $C_1=\Big((864JE_B^2 + 288KE_B^2 - 216J^3 - 216J^2K - 126JK^2 - 26K^3 + 6(-3072E_B^6 + (13824J^2 + 9216JK + 384K^2)E_B^4 + (-15552J^4 - 20736J^3K - 14688J^2K^2 - 5184JK^3 - 816K^4)E_B^2 - 27J^2K^4 - 18JK^5 - 9K^6)^{1/2}\Big)^{1/3}$.}
\centering
\begin{tabular}{|c|c|}
\hline
$|S_{tot},S_z\rangle$ & Eigenvalues \\ \hline

$|2,2\rangle$& $\frac{\mathit{C_1}}{6}+\frac{24 \mathit{E_B}^{2}+18 J^{2}+12 J K +5 K^{2}}{3 \mathit{C_1}}+\frac{J}{2}+\frac{K}{6}$ \\ \hline

$|2,1\rangle$& $\frac{3J}{2} + \frac{K}{2}+E_B$ \\ \hline

$|2,0\rangle$& $\frac{1}{2}\sqrt{9J^2 + 6JK + 3K^2}$ \\ \hline

$|2,-1\rangle$& $\frac{3J}{2} + \frac{K}{2}-E_B$ \\ \hline

$|2,-2\rangle$ ($|0,0\rangle$)& $3\Big( \left(\frac{1}{36}-\frac{\mathrm{I} \sqrt{3}}{36}\right) \mathit{C_1} - \left(\frac{J}{6}+\frac{K}{18}\right)$ \\

& $+ \left(\frac{4}{3} \mathit{E_B}^{2}+J^{2}+\frac{2JK}{3} +\frac{5K^2}{18}\right) \frac{\left(\mathrm{I} \sqrt{3}+1\right)}{C_1}\Big)
$ \\ \hline

$|1,S_z\rangle$& $-\frac{J}{2} - \frac{K}{2}+S_zE_B$ \\ \hline

$|1,1\rangle$& $-\frac{J}{2} + \frac{1}{2}\sqrt{4E_B^2 + K^2}$ \\ \hline

$|1,0\rangle$& $-\frac{J}{2} + \frac{K}{2}$ \\ \hline

$|1,-1\rangle$& $-\frac{J}{2} - \frac{1}{2}\sqrt{4E_B^2 + K^2}$ \\ \hline

$|1,1\rangle$& $-\frac{J}{2} + \frac{1}{2}\sqrt{4E_B^2 + K^2}$ \\ \hline

$|1,0\rangle$& $-\frac{J}{2} - \frac{K}{2}$ \\ \hline

$|1,-1\rangle$& $-\frac{J}{2} - \frac{1}{2}\sqrt{4E_B^2 + K^2}$ \\ \hline

$|0,0\rangle$ ($|2,-2\rangle$)& $3\Big( \left(-\frac{\mathrm{I} \sqrt{3}}{36}-\frac{1}{36}\right) \mathit{C_1} + \left(\frac{J}{6}+\frac{K}{18}\right)$ \\

& $+ \left(\frac{4}{3} \mathit{E_B}^{2}+J^{2}+\frac{2JK}{3} +\frac{5K^2}{18}\right) \frac{\left(\mathrm{I} \sqrt{3}-1\right)}{C_1}\Big)
$ \\ \hline

$|0,0\rangle$& $-\frac{1}{2}\sqrt{9J^2 + 6JK + 3K^2}$ \\ \hline

\end{tabular}%
\label{tab:symmetric_tetrahedron_eigenvalues}
\end{table}

\section{Tetrahedron}

Our final case, seen in Fig. \ref{fig:clusterfigures}, becomes much more complex, as the introduction of the second set of nearest neighbor (NN$_2$) interactions $J_2$ and $K_2$ combines both the coupled trimer and three-sided star tetramer configurations. For the tetrahedron model, the
Hamiltonian can be reduced to
\begin{equation}
\begin{array}{c}
H = J_1\Big(\vec{\sigma_1}\cdot\vec{\sigma_2} + \vec{\sigma_1}\cdot\vec{\sigma_3} +\vec{\sigma_1}\cdot\vec{\sigma_4}\Big)\\
+J_2\Big(\vec{\sigma_2}\cdot\vec{\sigma_3}+\vec{\sigma_2}\cdot\vec{\sigma_4}+\vec{\sigma_3}\cdot\vec{\sigma_4} \Big) \\  
+ K_1\Big( \sigma^{x}_1\cdot\sigma^{x}_2 + \sigma^{y}_1\cdot\sigma^{y}_3 + \sigma^{z}_1\cdot\sigma^{z}_4.\Big)\\
+K_2\Big( \sigma^{x}_3\cdot\sigma^{x}_4 + \sigma^{y}_2\cdot\sigma^{y}_4 + \sigma^{z}_2\cdot\sigma^{z}_3  \Big)
\end{array}
\end{equation}

Within the spin-1/2 basis, the tetrahedron model has the same spin decomposition and magnetic states as the three-sided star tetramer. Due to the presence of a second nearest-neighbor interaction, the energy levels become more degenerate with respect to the magnetic field, and the Kitaev NN$_2$ interactions do not affect the symmetry, but the Heisenberg NN$_2$ do \cite{Liu24IOP}. As the system grows in complexity, the number of eigenstates increases correspondingly, becoming too large to display explicitly.

\begin{table*}
\centering
\caption{The eigenstates $|S_{tot},S_z\rangle$ and eigenvalues for the asymmetric tetrahedron with no Kitaev interaction or no magnetic field.}
\begin{tabular}{|c|c|c|}
\hline
$|S_{tot},S_z\rangle$ & Eigenvalues & Eigenvalues\\
 & $K$=0 & $E_B$ = 0\\ \hline
 
$|2,S_z\rangle$  &  $\frac{3J_1}{4} + \frac{3J_2}{4}+S_zE_B$ & $\frac{J_1}{2} + \frac{1}{4}\sqrt{J_1^2 + 6J_1J_2 + 2J_1K_1 + 2J_1K_2 + 9J_2^2 + 6J_2K_1 + 6J_2K_2 + 3K_1^2 - 2K_1K_2 + 3K_2^2}$ \\

  &   & $\frac{1}{4}\sqrt{9J_1^2 + 18J_1J_2 + 6J_1K_1 + 6J_1K_2 + 9J_2^2 + 6J_2K_1 + 6J_2K_2 + 3K_1^2 + 6K_1K_2 + 3K_2^2}$ \\ \hline

$|1,S_z\rangle$  & $\frac{J_1}{4} - \frac{3J_2}{4}+S_zE_B$ & $\frac{J_1}{2} - \frac{1}{4}\sqrt{J_1^2 + 6J_1J_2 + 2J_1K_1 + 2J_1K_2 + 9J_2^2 + 6J_2K_1 + 6J_2K_2 + 3K_1^2 - 2K_1K_2 + 3K_2^2}$ \\ \hline

$|1,S_z\rangle$  & $\frac{J_1}{4} - \frac{3J_2}{4}+S_zE_B$ & $-\frac{J_1}{2} + \frac{1}{4}\sqrt{9J_1^2 - 18J_1J_2 + 10J_1K_1 - 6J_1K_2 + 9J_2^2 - 10J_2K_1 + 6J_2K_2 + 3K_1^2 - 2K_1K_2 + 3K_2^2}$ \\ \hline

$|1,S_z\rangle$  & $-\frac{5J_1}{4} + \frac{3J_2}{4}+S_zE_B$ & $-\frac{J_1}{2} - \frac{1}{4}\sqrt{9J_1^2 - 18J_1J_2 + 10J_1K_1 - 6J_1K_2 + 9J_2^2 - 10J_2K_1 + 6J_2K_2 + 3K_1^2 - 2K_1K_2 + 3K_2^2}$ \\ \hline

$|0,0\rangle$  & $-\frac{3J_1}{4} - \frac{3J_2}{4}$ & $-\frac{1}{4}\sqrt{9J_1^2 + 18J_1J_2 + 6J_1K_1 + 6J_1K_2 + 9J_2^2 + 6J_2K_1 + 6J_2K_2 + 3K_1^2 + 6K_1K_2 + 3K_2^2}$ \\ \hline

$|0,0\rangle$  & $-\frac{3J_1}{4} - \frac{3J_2}{4}$ & $-\frac{1}{4}\sqrt{9J_1^2 + 18J_1J_2 + 6J_1K_1 + 6J_1K_2 + 9J_2^2 + 6J_2K_1 + 6J_2K_2 + 3K_1^2 + 6K_1K_2 + 3K_2^2}$ \\ \hline

\end{tabular}%
\label{tab:asymmetric_tetrahedron_eigenvalues}
\end{table*}

\subsection{Symmetric tetrahedon}

\begin{figure*}
    \centering
    \includegraphics[width= 6 in]{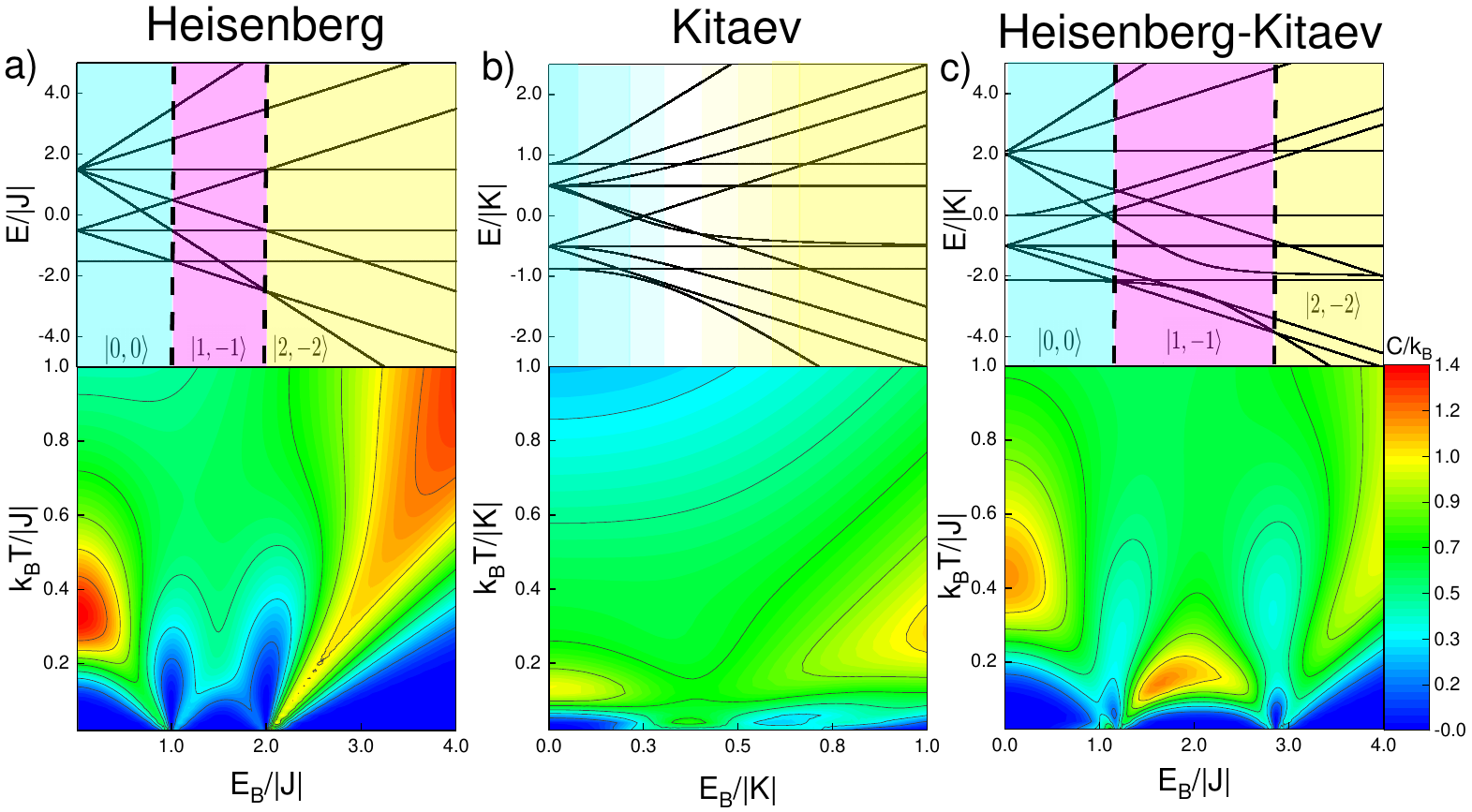}
    \caption{
    The energy eigenvalues with respect to the magnetic field and their corresponding temperature-dependent heat capacity for the symmetric tetrahedron. In the Heisenberg case (a), the peak in the heat capacity associated with $NN_1$ precedes two first-order phase transitions from $S_{tot} = 0$ to $S_{tot} = 1$ and from $S_{tot} = 1$ to $S_{tot} = 2$  and a peak associated with $NN_2$ appears after the transition under the influence of a magnetic field. In contrast, the Kitaev case (b) exhibits a field-driven second-order phase transition directly from $S_{tot} = 0$ to $S_{tot} = 2$. The first-order transitions observed in (a) also appear in the Heisenberg-Kitaev model (c), though a new peak appears between the two transitions.
    }
    \label{fig:SymmTetrahedroninteractions}
\end{figure*}

The symmetric tetrahedron produces eigenvalues from the full Hamiltonian in table \ref{tab:symmetric_tetrahedron_eigenvalues}. Starting with $J_1 = J_2$, and $K_1 = K_2 = 0$ the energy eigenstates and heat capacity are shown in Fig. \ref{fig:SymmTetrahedroninteractions}. Due to symmetry, the energy levels split into only nine magnetic sub-states. Compared to the previous cases, the initial splitting exhibits behavior similar to the Heisenberg trimer. Interestingly, this causes the higher-order transitions observed in the tetramer case to become the first ground-state transition despite significantly less spin mixing. The ground state begins as an $S_{tot} = 0$ state, but at $\frac{E_B}{|J|} = 1.0$, the system transitions into an $S_{tot} = 1$ state. With further increase in the field, the system undergoes a second phase shift at $\frac{E_B}{|J|} = 2.0$, reaching an $S_{tot} = 2$ state. In relation to the heat capacity, the peak associated with the Heisenberg NN$_1$ interaction is observed at low field from $\frac{K_B T}{|J|}$ = 0.2 to 0.6. As the NN$_1$ interactions dominate, the behavior seen from $\frac{E_B}{|J|}$ = 0 to 1.0 is qualitatively similar to a shifted three-sided star tetramer. Following the quantum phase transition, the system simultaneously shifts to being NN$_2$-driven, exhibiting behavior similar to that of the trimer case. Likewise, a Schottky anomaly appears after the transition from $S_{tot} = 1$ to $S_{tot} = 2$.

Now setting $J_1 = J_2 = 0.0$ and $K_1 = K_2$, the Kitaev tetrahedron produces many more initial energy levels than the Kitaev tetramer, but the shape is similar. However, the amount of field required to cause the same spin mixing is decreased. The system's ground state remains $S_{tot} = 0$ until $\frac{E_B}{|K|} \approx 0.05$, where the onset of nonlinearity becomes apparent, and the ground state begins to evolve with the field. Instead of a distinct quantum phase transition, the ground state approaches other $S_{tot} = 1$ eigenstates without becoming exactly degenerate before gradually curving into an $S_{tot} = 2$ ground state. The heat capacity reflects the absence of a distinct transition, capturing the gradual curvature of the ground state energy, which produces a peak below $T = 0.2$ at low field. As the field increases, this peak diminishes, but a small secondary peak emerges around $\frac{E_B}{|K|} = 0.4$. With further increases in the field, exotic behavior appears, followed by the emergence of the familiar broad peak.

With $J_1 = J_2 = K_1 = K_2 = 1.0$, Heisenberg behavior primarily governs energy levels. However, level repulsion between two eigenstates introduces nonlinear behavior, suppressing a higher-order crossover feature. Similarly, some of the eigenstates exhibit nonlinear behavior initially. Still, around $\frac{E_B}{|J|} = 0.5$, they transition to and maintain linear behavior as the J-K parameter coupling becomes overcome by the field. The enhanced symmetry of the tetrahedron introduces more significant degeneracy in the system. However, the Kitaev interaction breaks some of this intrinsic symmetry, lifting certain degeneracies and leading to a more complex energy spectrum. 

Several energy levels appear to align or remain close to each other at certain points, indicating partial degeneracy. As the magnetic field increases, some initial degenerate levels are lifted, breaking symmetry. The ground-state energy initially varies linearly in the $S_{tot} = 0$ state, transitioning to the $S_{tot} = 1$ state at $\frac{E_B}{|J|} \approx 1.2$. The original lowest-energy eigenstate lifts its degeneracy as it becomes nonlinear, curving with the field until it crosses again at $\frac{E_B}{|J|} \approx 2.8$, followed by a final field-driven transition to the $S_{tot} = 2$ state. Evaluating the impact on heat capacity, we demonstrate that the effects of the ground state transition are constrained, revealing a new feature. Between the transitions, a peak in the heat capacity emerges due to both $K_1$ and $K_2$, representing a distinct feature associated with partial degeneracies between linear and nonlinear eigenstates. In this arrangement, the second quantum phase transition requires a stronger field to occur, and the sharpness of the Schottky anomaly introduced by the $J_2$ interaction is significantly reduced.

\subsection{Asymmetric tetrahedron}

\begin{figure*}
    \centering
    \includegraphics[width= 6 in]{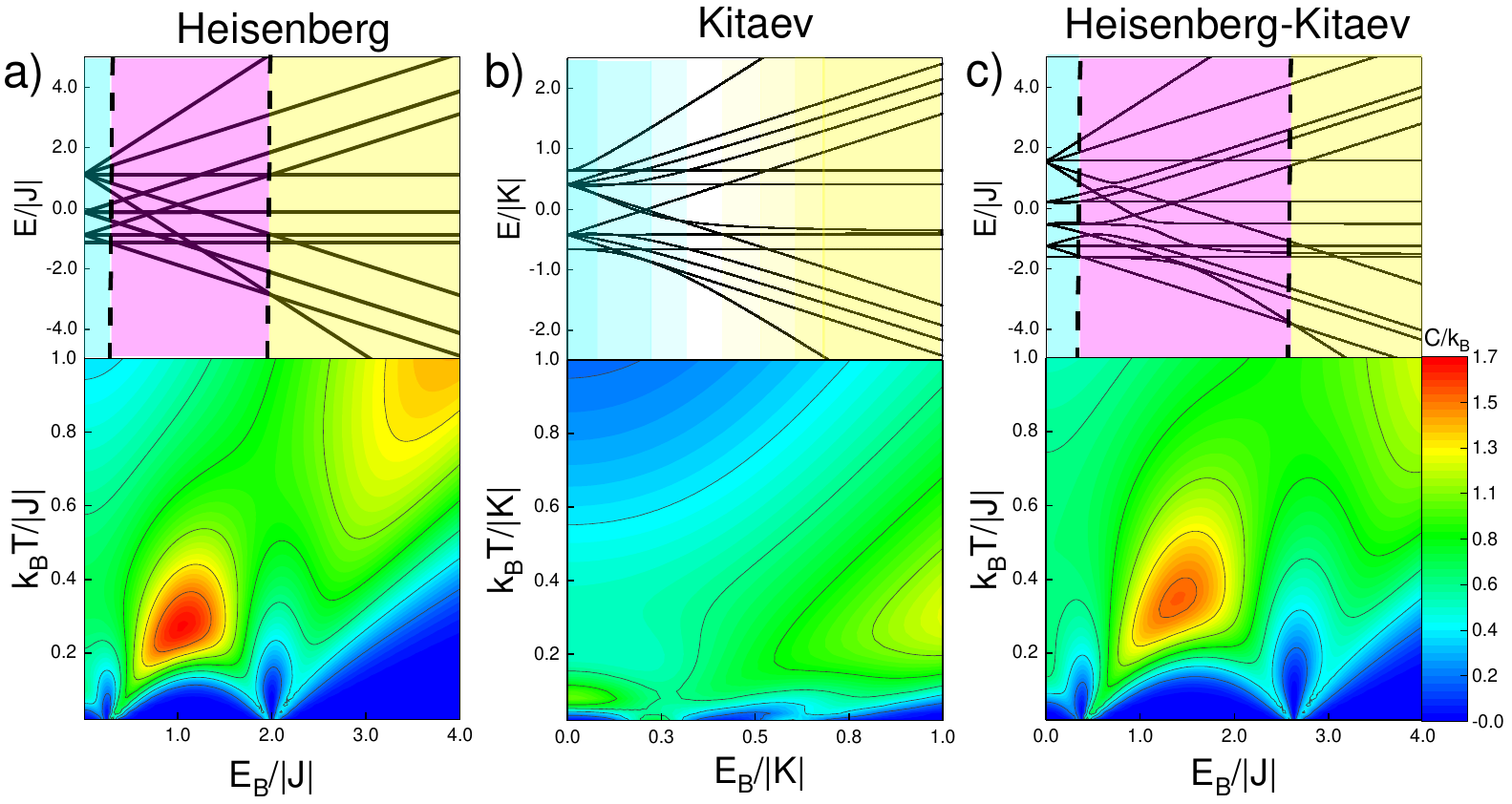}
    \caption{The energy eigenvalues with respect to the magnetic field and their corresponding temperature-dependent heat capacity for the $NN_1$ dominated tetrahedron. In the Heisenberg case (a), first-order phase transitions from $S_{tot} = 0$ to $S_{tot} = 1$ and from $S_{tot} = 1$ to $S_{tot} = 2$ emerge under the influence of a magnetic field. In contrast, the Kitaev case (b) exhibits a field-driven second-order phase transition directly from $S_{tot} = 0$ to $S_{tot} = 2$. The first-order transitions observed in (a) also appear in the Heisenberg-Kitaev model (c), though they require a stronger magnetic field to be induced. In case (a), the peak in the heat capacity associated with $NN_1$ is observed. In case (c), the second phase transition is strongly altered.}
    \label{fig:NN1AsymmTetrahedroninteractions}
\end{figure*}

The eigenvalues for the asymmetric tetrahedron where $K=0$ and $E_B = 0$ respectively, are provided in Table \ref{tab:asymmetric_tetrahedron_eigenvalues}.  Using the previous cases as a reference, the $NN_1$ and $NN_2$ interactions for the isotropic and anisotropic exchange terms are assigned different strengths to investigate the influence of each interaction and the impact of asymmetry. We begin by considering the interactions associated with the three-sided star tetramer, with the $NN_1$ interaction dominating the system in Fig. \ref{fig:NN1AsymmTetrahedroninteractions} a). By only considering the Heisenberg interactions, it is observed that the system spends less time in the $|0,0\rangle$ state and requires less magnetic field to produce a first-order phase transition to $|1,-1\rangle$. By comparing the energy eigenvalues with the corresponding temperature-dependent heat capacity, we observe that the peak appearing on the left side of the first transition in the symmetric tetrahedron, as shown in Fig. \ref{fig:SymmTetrahedroninteractions} a), is shifted to lie between the two phase transitions. This indicates that the $NN_1$ interactions predominantly drive the first quantum phase transition, and as their influence intensifies, the magnetic field required to induce the first transition decreases. In contrast, the second phase transition following the peak requires the same field strength but spans a narrower temperature range as the mixed spin states begin to dominate the system. Furthermore, the shoulder corresponding to the coupled trimer interactions $NN_2$ after the second transition becomes less pronounced.

Next, we consider anisotropic interactions, which reveal that the effects of the asymmetric tetrahedron, with nearest-neighbor $NN_1$ interactions dominating, are less pronounced compared to isotropic interactions. The second-order transition associated with the symmetric tetrahedron persists, but there is significantly more spin mixing in the vicinity of the $|1,-1\rangle$ state. The impact of asymmetry on the heat capacity is evident as the smaller peaks decrease in intensity but become more distinctly separated as fewer thermal states are populated at higher temperatures in Fig. \ref{fig:NN1AsymmTetrahedroninteractions} b). As the isotropic and anisotropic interactions are combined, the effects of the interplay of the magnetic interactions are shown in Fig. \ref{fig:NN1AsymmTetrahedroninteractions} c). The inclusion of the Kitaev interaction lifts all remaining degenerate states, allowing the two spin transitions to extend into higher temperature ranges and increasing the magnetic field required to induce spin transitions. As the system begins to favor the $NN_1$ configuration, the inherent frustration is reduced, resulting in a single peak rather than three.

\begin{figure*}
    \centering
    \includegraphics[width= 6 in]{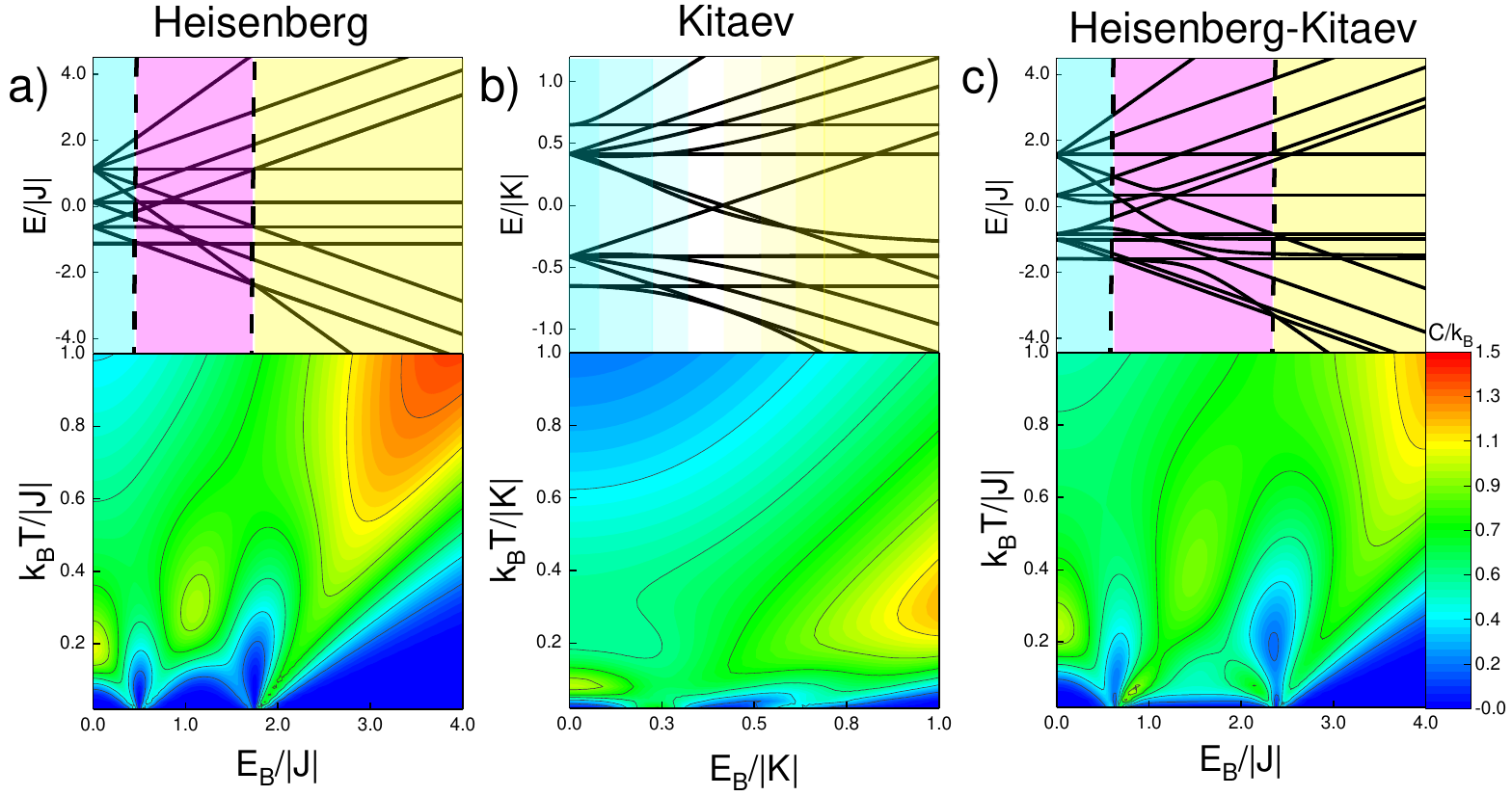}
    \caption{The energy eigenvalues with respect to the magnetic field and their corresponding temperature-dependent heat capacity for the $NN_2$ dominated tetrahedron. In the Heisenberg case (a), first-order phase transitions from $S_{tot} = 0$ to $S_{tot} = 1$ and from $S_{tot} = 1$ to $S_{tot} = 2$ emerge under the influence of a magnetic field. In contrast, the Kitaev case (b) exhibits a field-driven second-order phase transition directly from $S_{tot} = 0$ to $S_{tot} = 2$. The first-order transitions observed in (a) also appear in the Heisenberg-Kitaev model (c), though they require a stronger magnetic field to be induced. In case (a), the second peak in the heat capacity associated with $NN_1$ is observed, along with a new peak that precedes the first phase transition. A prominent peak corresponding to $NN_2$ is also present. In case (c), the heat capacity enters a frustrated regime between the two phase transitions, diminishing $NN_1$ contributions and populating new thermal states at lower temperatures. }
    \label{fig:NN2ASymmTetrahedroninteractions}
\end{figure*}

Another form of asymmetry arises when the tetrahedron exhibits stronger $NN_2$ interactions within the configuration. Similarly to the previous case, we start with isotropic interactions associated with the coupled trimer, where $NN_2$ dominates the system, as shown in Fig. \ref{fig:NN2ASymmTetrahedroninteractions} a). In this situation, the time spent in the $|0,0\rangle$ state decreases compared to the symmetric configuration but increases relative to the $NN_1$ dominated configuration. A small peak appears before the first spin transition, while the peak from the $NN_1$-dominated configuration occurs between the two transitions, albeit within a narrower temperature range and with a lower heat capacity. As the $NN_2$ interaction increases, the field required to induce the phase transition from $|1,-1\rangle$ to $|2,-2\rangle$ decreases, while the associated thermodynamic effects extend to higher temperature ranges. The second-order phase transition linked to the Kitaev interaction still occurs in Fig. \ref{fig:NN2ASymmTetrahedroninteractions}, but with reduced degeneracy and more crossover points, resulting in an intensified rightmost peak. The interplay between the Heisenberg and Kitaev interactions within this asymmetric configuration results in significant changes in the heat capacity, driven by energy levels that are in close proximity but not degenerate. This proximity allows more quantum states to be populated at lower temperatures. The intermediate peak that appeared between transitions in Fig. \ref{fig:NN2ASymmTetrahedroninteractions} a) is diminished, while two smaller peaks emerge near each transition, reflecting a redistribution of spectral weight as the system transitions between magnetic configurations. The transition from lower to higher temperature ranges reflects the changing influence of magnetic field strength on spin configurations, with the sharper peak indicating reduced degeneracy compared to the energy levels. The regime in which the system is in the $|1,-1\rangle$ state exhibits the most robust and exotic thermodynamic behavior at lower temperatures, reducing the magnetic field's influence on the system's Kitaev components and resulting in a highly frustrated state.

\section{Implications to Experiments}

The ability to experimentally realize these interactions is crucial to understanding and exploring their potential applications. Therefore, we now focus on discussing some of the experimental implications of these interactions for inelastic neutron scattering and magnetic heat capacity, as well as highlighting some of the signatures that experimentalists can look for in these types of systems.

\subsection{Inelastic Neutron Scattering}

Inelastic neutron scattering is a valuable tool for examining magnetic interactions, as it probes systems at the microscopic level. It is well known that neutron scattering excites magnetic transitions of $\pm$1 and 0 in total spin\cite{Squire78TNS}. However, the $S_z$ component and the geometry can also play a role in restricting neutron scattering excitations \cite{Haraldsen11PRL}. Therefore, when different systems undergo a quantum phase transition with an increasing magnetic field, there will typically be a corresponding change in the number of excitations observed, especially when transitioning to a higher spin state. 

For the trimer, the ground state will go from a spin 1/2 to a spin 3/2. At zero field, there will be six total excitations from $|1/2,S_z\rangle$ to $|3/2,S_z\rangle$ that are all degenerate in energy. However, once the Zeeman splitting occurs, this will drop to four transitions, where one is the $|1/2,1/2\rangle$ to $|1/2,-1/2\rangle$ transition and three are $|1/2,1/2\rangle$ to $|3/2,S_z\rangle$ ($S_z$ = 3/2,1/2, and -1/2). Once the trimer transitions to the $|3/2,-3/2\rangle$ state, there is only one possible transitions ($|3/2,-3/2\rangle$ to $|1/2,-1/2\rangle$). 

While these transitions do not seem like they will change in either the Heisenberg or Kitaev pictures, Figure \ref{fig:CoupledTrimerModel} shows the spin mixing between the $|3/2,-3/2\rangle$ and one of the $|1/2,1/2\rangle$ levels. The spin mixing does not change the overall ground state, in this case. Still, it splits the degeneracy of the $|1/2,-1/2\rangle$ energy levels and creates additional transitions that can be observed in inelastic neutron scattering.

The spin mixing has a similar effect on the tetramer transitions. As shown in Figs. \ref{fig:tetramerinteractions} and \ref{fig:SymmTetrahedroninteractions}, the addition of the Kitaev splits the $S_z$ = 0 levels and produces spin mixing between the $|2,-2\rangle$ and $|0,0\rangle$ states. Therefore, the presence of Kitaev interactions can be determined by an examination of the energy excitations by inelastic neutron scattering.

\begin{figure*}
    \centering
    \includegraphics[width=5in]{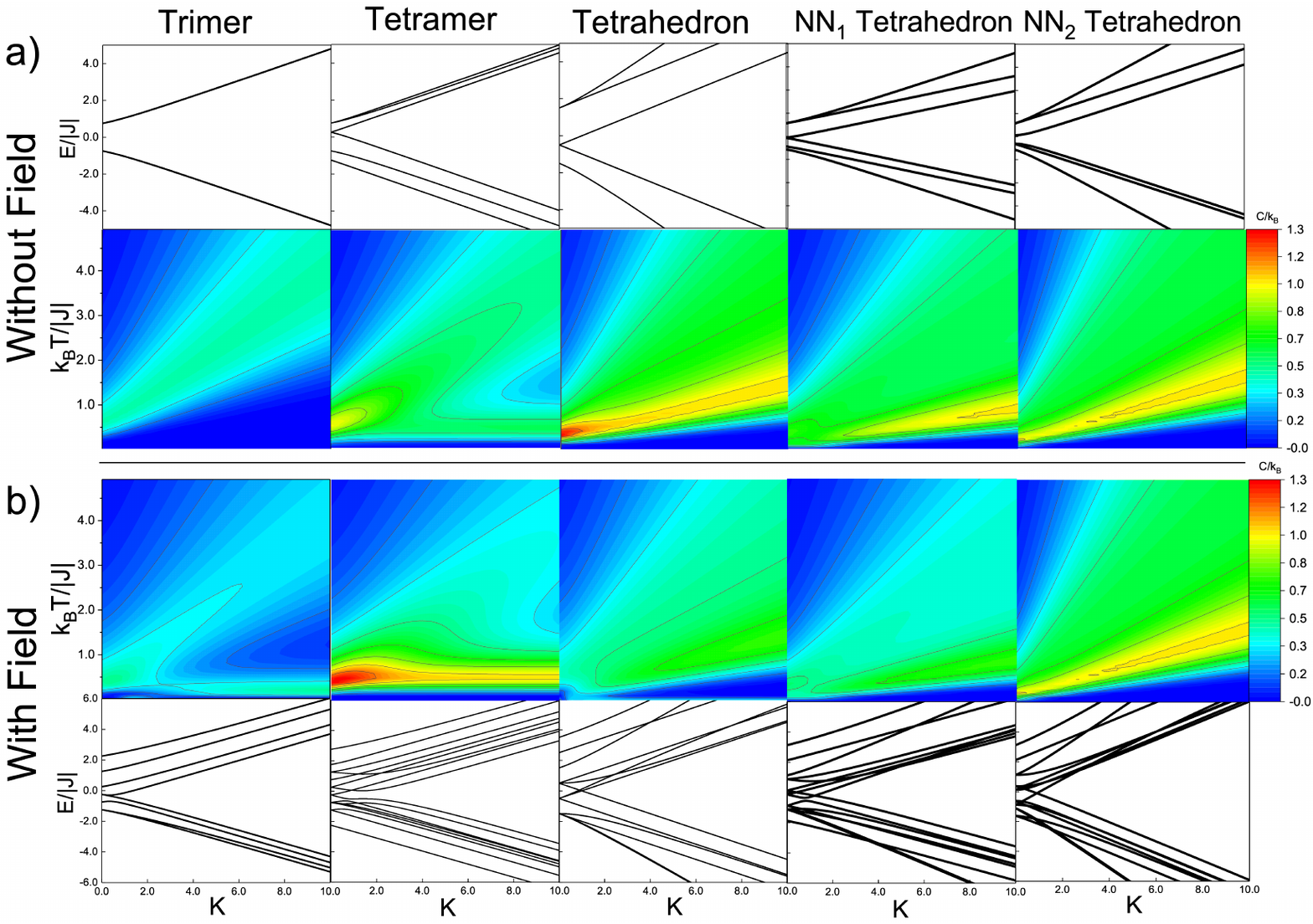}
    \caption{Energy eigenvalues are shown as a function of the Kitaev term, corresponding to their temperature-dependent Kitaev signatures in the heat capacity: (a) without an applied magnetic field and (b) with an applied magnetic field, for the spin-1/2 trimer, tetramer, symmetric tetrahedron and asymmetric tetrahedron clusters. }
    \label{fig:KitaevSignature}
\end{figure*}

\subsection{Magnetic Heat Capacity}

In order to record Kitaev-specific signatures within these spin arrangements, we set the Heisenberg interaction $J$ = 1.0 and plot the energy eigenvalues and temperature-dependent heat capacity with respect to $\frac{K}{|J|}$, which shows how the non-perturbed systems go from a Heisenberg dominated regime to a Kitaev dominated system, the effects on the energy levels and heat capacity are observed in Fig. \ref{fig:KitaevSignature} a). For the H-K trimer, as the Kitaev interaction increases a singular Schottky anomaly appears, despite the existences of two order parameters. The nonlinearity due to the coupling between $J-K$ is subtly observable in the energy levels. 

The H-K tetramer in a three-sided star configuration exhibits energy state splitting as $K$ increases. There are nonlinear effects in two of the seven degenerate energy levels at low $K/J$, but they vanish in all but the highest excited state. The heat capacity exhibits two distinct Schottky anomalies: the diagonal anomaly reflects the Kitaev contributions, similar to the trimer case, while the horizontal anomaly arises from the Heisenberg parameter. The interplay between the two parameters becomes pronounced due to the central spin, which acts as a conduit, facilitating enhanced exchange interactions across the system. 

Combining the two configurations produces a Heisenberg-Kitaev tetrahedron that gives rise to two levels of frustration. The first is coupling the $J$ and $K$ parameters, and the second is the interrelation between the NN$_1$ and NN$_2$ interactions. The competition of exchanges highlights the nonlinearity introduced by the coupling of parameters, which does not result from level repulsion. More importantly, only the highest excited state displays the curved behavior. Compared to the previous two cases, the Kitaev contribution becomes obscured as the two Schottky anomalies converge and the effects spread through higher temperatures and lower $K/J$ values.

Furthermore, in Fig. \ref{fig:KitaevSignature} b), we observe the response of the thermodynamic signatures of the Kitaev interaction to an applied magnetic field. The trimer energy levels split into six magnetic sub-levels, which become increasingly nonlinear as $K$ increases. Notably, the addition of a magnetic field confirms its role in causing level repulsion, now observed between two eigenstates where no such repulsion existed previously. Seemingly, this effect on the Kitaev-dependent heat capacity results in the emergence of a second Schottky anomaly, while the original anomaly diminishes in both span and magnitude relative to the magnetic field. This secondary low-temperature peak is attributed to decoupling the Heisenberg and Kitaev terms. 

Spin mixing from the central spin creates an entangled state within the H-K tetramer in a three-sided star configuration. The six energy eigenstates, initially split by $ K/J $, are further divided under an applied field, producing 16 distinct eigenstates, with 6 displaying level repulsion. This repulsion and the curvature introduced by $ K/J $ drive multiple higher-order spin transitions as $ K/J $ progresses from 1 to 2. The resultant spin entanglement interactions in this range produce a sharp peak in the heat capacity. As the Kitaev exchange begins to dominate the interaction ratio, the resulting eigenstates approach each other instead of diverging, causing the Schottky anomaly associated with the $ K $ parameter to merge with the Schottky anomaly linked to the $ J $ parameter.

Consequently, as the two configurations merge into the H-K tetrahedron with nearest-neighbor (NN$_1$) tetramer and next-nearest-neighbor (NN$_2$) trimer interactions, the addition of asymmetry under the applied field $ E_B = 1.0 $ increases the number of spin coupling effects and increases the number of degenerate states, resulting in 12 distinct behaviors. As $ K $ increases, immediate competition between the interactions gives rise to nonlinear behavior, which causes degenerate eigenstates to diverge, and at $ K/J = 1 $, level repulsion begins just before the first linear higher-order spin transition. At this point, the Kitaev, Heisenberg, and Zeeman terms have equal contributions, and degeneracies partially break, involving three eigenstates in the transition. The reduction in degeneracy and the overall number of crossings in this range eliminate the sharp peak observed in the tetramer case. Instead, the competition between the NN$_1$ and NN$_2$ interactions produces a very broad, singular Schottky anomaly in the heat capacity. Compared to the results without a field, the peak associated with the competition between $ J $-$ K $-$E_B$ couplings vanish, and the overall heat capacity values decrease as more spin states contribute to the heat capacity.

\section{Conclusion }

We identified several observations regarding the behavior of the trimer, tetramer, and tetrahedron spin clusters, systems that have been of interest experimentally in Refs.~\cite{Drulis1985, Kornjaca2023}. By adding a Kitaev interaction to the spin clusters, exact eigenvalues and thermodynamic signatures are determined for spin geometries associated with molecular magnets and Kitaev spin liquids. The Kitaev interaction demonstrates high sensitivity to external magnetic fields, particularly within the range where $\frac{E_B}{|K|}$ $<$ 1. In this range, the energy levels split, and level repulsion emerges. Although Kitaev contributions have not been explored experimentally in molecular magnets, these repulsion effects have been observed in spin orbit-coupled molecular magnets~\cite{Plumb2016NatPhy,  Shivaram_2017, Kragskow2022}, as well as in heat capacity and electron spin resonance experiments on SOC materials~\cite{Liu2018,  Vaclavkova2021}, including good agreement with proximate QSL, where long-range magnetic order is suppressed \cite{Ponomaryov2017, Sichelschmidt2019, Li2021}. 

A first-order quantum phase transition is observed at low magnetic fields in the Kitaev trimer and tetramer, driven by energy level repulsion. However, the pronounced behavior seen in the trimer's heat capacity is reduced in the tetramer due to spin mixing. In the Kitaev tetrahedron, the transition is absent due to competing interactions between nearest neighbors ($K_1$) and second-nearest neighbors ($K_2$). Consequently, the heat capacity assumes a shape reminiscent of the Heisenberg case, but at lower magnetic fields.

For Heisenberg-Kitaev spin clusters, isotropic interactions dictate the overall shape of the temperature-dependent heat capacity in a magnetic field, while anisotropic terms refine its details, leading to exotic low-temperature behavior near quantum phase transitions. This effect is most pronounced in the trimer and tetrahedron. In both the trimer and tetramer, the Kitaev term increases the field strength required for a ground-state transition.

Through asymmetric variations, we find that the first quantum phase transition in the tetrahedron stems from isotropic nearest-neighbor interactions, while the second is driven by next-nearest-neighbor interactions. Competition between isotropic and anisotropic terms is strongest in the tetrahedron, where NN$_1$–NN$_2$ frustration creates a new peak between transitions. The Kitaev interaction significantly affects NN$_2$ contributions, increasing the field required for a transition and narrowing the temperature range over which these effects occur. While NN$_1$ interactions remain notable, spin mixing diminishes the Kitaev term’s overall impact.

We demonstrate the thermodynamic signatures of the Kitaev interaction across all three configurations. The trimer, due to its symmetry, exhibits a single Schottky anomaly despite having two order parameters. A magnetic field lifts degeneracies, splitting it into two anomalies with reduced heat capacity. In the tetramer, two distinct anomalies appear, but the field disrupts the Kitaev-related anomaly, merging it with the Heisenberg anomaly. In the tetrahedron, frustration between interactions further obscures a clear Schottky anomaly, an effect intensified by an applied field. Ultimately, the sensitivity of the Kitaev interaction to magnetic fields disrupts its signature within the Heisenberg-Kitaev framework.

\section{Acknowledgments}

The authors acknowledge support from the Institute for Materials Science at Los Alamos National Laboratory and support from the Materials Science and Engineering Program at the University of North Florida. Work at Los Alamos was carried out under the auspices of the U.S. Department of Energy (DOE) National Nuclear Security Administration (NNSA) under Contract No. 89233218CNA000001 and LA-UR-25-25164. It was supported by Quantum Science Center, a U.S. DOE Office of Science Quantum Information Science Research Center, and in part by Center for Integrated Nanotechnologies, a DOE BES user facility, in partnership with the LANL Institutional Computing Program for computational resources.

\section{Appendix A}

The matrices \ref{eq:wideeq} and \ref{eq:tetrahedron} represent the Hamiltonians for tetramer and tetrahedron spin clusters with Heisenberg-Kitaev (H-K) interactions under an applied magnetic field. The 16x16 tetramer Hamiltonian matrix incorporates Heisenberg ($ J $) and Kitaev ($ K $) terms and the magnetic field term $ E_B $, where off-diagonal elements capture coupling between states. These interactions, especially from the Kitaev terms, introduce anisotropic coupling and level repulsion effects, leading to energy level splitting. The tetrahedron matrix introduces an additional level of complexity, as it incorporates distinct parameters $ J1 $, $ J2 $, $ K1 $, and $ K2 $ to allow for asymmetrical exchange interactions among the four spins. Diagonalizing both matrices provides energy eigenstates that reveal key thermodynamic signatures of the Kitaev interaction and the system's response to magnetic fields, with observable effects such as level repulsion and quantum phase transitions under changing field strengths.

\begin{widetext}
\begin{equation}
\resizebox{\textwidth}{!}{$
H = \begin{pmatrix}
\begin{array}{cccccccccccccccc}
\frac{3 J}{4}+\frac{K}{4}+2 \mathit{E_B} & 0 & 0 & 0 & 0 & 0 & 0 & 0 & 0 & 0 & -\frac{K}{4} & 0 & \frac{K}{4} & 0 & 0 & 0 \\
0 & \frac{J}{4}-\frac{K}{4}+\mathit{E_B} & 0 & 0 & 0 & 0 & 0 & 0 & \frac{J}{2} & 0 & 0 & -\frac{K}{4} & 0 & \frac{K}{4} & 0 & 0 \\
0 & 0 & \frac{J}{4}+\frac{K}{4}+\mathit{E_B} & 0 & 0 & 0 & 0 & 0 & \frac{J}{2}+\frac{K}{4} & 0 & 0 & 0 & 0 & 0 & \frac{K}{4} & 0 \\
0 & 0 & 0 & -\frac{J}{4}-\frac{K}{4} & 0 & 0 & 0 & 0 & 0 & \frac{J}{2}+\frac{K}{4} & \frac{J}{2} & 0 & 0 & 0 & 0 & \frac{K}{4} \\
0 & 0 & 0 & 0 & \frac{J}{4}+\frac{K}{4}+\mathit{E_B} & 0 & 0 & 0 & \frac{J}{2}+\frac{K}{4} & 0 & 0 & 0 & 0 & 0 & -\frac{K}{4} & 0 \\
0 & 0 & 0 & 0 & 0 & -\frac{J}{4}-\frac{K}{4} & 0 & 0 & 0 & \frac{J}{2}+\frac{K}{4} & 0 & 0 & \frac{J}{2} & 0 & 0 & -\frac{K}{4} \\
0 & 0 & 0 & 0 & 0 & 0 & -\frac{J}{4}+\frac{K}{4} & 0 & 0 & 0 & \frac{J}{2}+\frac{K}{4} & 0 & \frac{J}{2}+\frac{K}{4} & 0 & 0 & 0 \\
0 & 0 & 0 & 0 & 0 & 0 & 0 & -\frac{3 J}{4}-\frac{K}{4}-\mathit{E_B} & 0 & 0 & 0 & \frac{J}{2}+\frac{K}{4} & 0 & \frac{J}{2}+\frac{K}{4} & \frac{J}{2} & 0 \\
0 & \frac{J}{2} & \frac{J}{2}+\frac{K}{4} & 0 & \frac{J}{2}+\frac{K}{4} & 0 & 0 & 0 & -\frac{3 J}{4}-\frac{K}{4}+\mathit{E_B} & 0 & 0 & 0 & 0 & 0 & 0 & 0 \\
0 & 0 & 0 & \frac{J}{2}+\frac{K}{4} & 0 & \frac{J}{2}+\frac{K}{4} & 0 & 0 & 0 & -\frac{J}{4}+\frac{K}{4} & 0 & 0 & 0 & 0 & 0 & 0 \\
-\frac{K}{4} & 0 & 0 & \frac{J}{2} & 0 & 0 & \frac{J}{2}+\frac{K}{4} & 0 & 0 & 0 & -\frac{J}{4}-\frac{K}{4} & 0 & 0 & 0 & 0 & 0 \\
0 & -\frac{K}{4} & 0 & 0 & 0 & 0 & 0 & \frac{J}{2}+\frac{K}{4} & 0 & 0 & 0 & \frac{J}{4}+\frac{K}{4}-\mathit{E_B} & 0 & 0 & 0 & 0 \\
\frac{K}{4} & 0 & 0 & 0 & 0 & \frac{J}{2} & \frac{J}{2}+\frac{K}{4} & 0 & 0 & 0 & 0 & 0 & -\frac{J}{4}-\frac{K}{4} & 0 & 0 & 0 \\
0 & \frac{K}{4} & 0 & 0 & 0 & 0 & 0 & \frac{J}{2}+\frac{K}{4} & 0 & 0 & 0 & 0 & 0 & \frac{J}{4}+\frac{K}{4}-\mathit{E_B} & 0 & 0 \\
0 & 0 & \frac{K}{4} & 0 & -\frac{K}{4} & 0 & 0 & \frac{J}{2} & 0 & 0 & 0 & 0 & 0 & 0 & \frac{J}{4}-\frac{K}{4}-\mathit{E_B} & 0 \\
0 & 0 & 0 & \frac{K}{4} & 0 & -\frac{K}{4} & 0 & 0 & 0 & 0 & 0 & 0 & 0 & 0 & 0 & \frac{3 J}{4}+\frac{K}{4}-2 \mathit{E_B} \\
\end{array}
\end{pmatrix}
$}
 \label{eq:wideeq}
\end{equation}
\end{widetext}

\begin{widetext}
\begin{equation}
\resizebox{\textwidth}{!}{$
H = \begin{pmatrix}
\begin{array}{cccccccccccccccc}
\frac{3 \mathit{J1}}{4}+\frac{3 \mathit{J2}}{4}+\frac{\mathit{K1}}{4}+\frac{\mathit{K2}}{4}+2 \mathit{E_B}  & 0 & 0 & \frac{\mathit{K2}}{4} & 0 & -\frac{\mathit{K2}}{4} & 0 & 0 & 0 & 0 & -\frac{\mathit{K1}}{4} & 0 & \frac{\mathit{K1}}{4} & 0 & 0 & 0 
\\
 0 & \frac{\mathit{J1}}{4}-\frac{\mathit{J2}}{4}-\frac{\mathit{K1}}{4}+\frac{\mathit{K2}}{4}+\mathit{E_B}  & \frac{\mathit{J2}}{2}+\frac{\mathit{K2}}{4} & 0 & \frac{\mathit{J2}}{2}+\frac{\mathit{K2}}{4} & 0 & 0 & 0 & \frac{\mathit{J1}}{2} & 0 & 0 & -\frac{\mathit{K1}}{4} & 0 & \frac{\mathit{K1}}{4} & 0 & 0 
\\
 0 & \frac{\mathit{J2}}{2}+\frac{\mathit{K2}}{4} & \frac{\mathit{J1}}{4}-\frac{\mathit{J2}}{4}+\frac{\mathit{K1}}{4}-\frac{\mathit{K2}}{4}+\mathit{E_B}  & 0 & \frac{\mathit{J2}}{2} & 0 & 0 & -\frac{\mathit{K2}}{4} & \frac{\mathit{J1}}{2}+\frac{\mathit{K1}}{4} & 0 & 0 & 0 & 0 & 0 & \frac{\mathit{K1}}{4} & 0 
\\
 \frac{\mathit{K2}}{4} & 0 & 0 & -\frac{\mathit{J1}}{4}-\frac{\mathit{J2}}{4}-\frac{\mathit{K1}}{4}-\frac{\mathit{K2}}{4} & 0 & \frac{\mathit{J2}}{2} & \frac{\mathit{J2}}{2}+\frac{\mathit{K2}}{4} & 0 & 0 & \frac{\mathit{J1}}{2}+\frac{\mathit{K1}}{4} & \frac{\mathit{J1}}{2} & 0 & 0 & 0 & 0 & \frac{\mathit{K1}}{4} 
\\
 0 & \frac{\mathit{J2}}{2}+\frac{\mathit{K2}}{4} & \frac{\mathit{J2}}{2} & 0 & \frac{\mathit{J1}}{4}-\frac{\mathit{J2}}{4}+\frac{\mathit{K1}}{4}-\frac{\mathit{K2}}{4}+\mathit{E_B}  & 0 & 0 & \frac{\mathit{K2}}{4} & \frac{\mathit{J1}}{2}+\frac{\mathit{K1}}{4} & 0 & 0 & 0 & 0 & 0 & -\frac{\mathit{K1}}{4} & 0 
\\
 -\frac{\mathit{K2}}{4} & 0 & 0 & \frac{\mathit{J2}}{2} & 0 & -\frac{\mathit{J1}}{4}-\frac{\mathit{J2}}{4}-\frac{\mathit{K1}}{4}-\frac{\mathit{K2}}{4} & \frac{\mathit{J2}}{2}+\frac{\mathit{K2}}{4} & 0 & 0 & \frac{\mathit{J1}}{2}+\frac{\mathit{K1}}{4} & 0 & 0 & \frac{\mathit{J1}}{2} & 0 & 0 & -\frac{\mathit{K1}}{4} 
\\
 0 & 0 & 0 & \frac{\mathit{J2}}{2}+\frac{\mathit{K2}}{4} & 0 & \frac{\mathit{J2}}{2}+\frac{\mathit{K2}}{4} & -\frac{\mathit{J1}}{4}-\frac{\mathit{J2}}{4}+\frac{\mathit{K1}}{4}+\frac{\mathit{K2}}{4} & 0 & 0 & 0 & \frac{\mathit{J1}}{2}+\frac{\mathit{K1}}{4} & 0 & \frac{\mathit{J1}}{2}+\frac{\mathit{K1}}{4} & 0 & 0 & 0 
\\
 0 & 0 & -\frac{\mathit{K2}}{4} & 0 & \frac{\mathit{K2}}{4} & 0 & 0 & -\frac{3 \mathit{J1}}{4}+\frac{3 \mathit{J2}}{4}-\frac{\mathit{K1}}{4}+\frac{\mathit{K2}}{4}-\mathit{E_B}  & 0 & 0 & 0 & \frac{\mathit{J1}}{2}+\frac{\mathit{K1}}{4} & 0 & \frac{\mathit{J1}}{2}+\frac{\mathit{K1}}{4} & \frac{\mathit{J1}}{2} & 0 
\\
 0 & \frac{\mathit{J1}}{2} & \frac{\mathit{J1}}{2}+\frac{\mathit{K1}}{4} & 0 & \frac{\mathit{J1}}{2}+\frac{\mathit{K1}}{4} & 0 & 0 & 0 & -\frac{3 \mathit{J1}}{4}+\frac{3 \mathit{J2}}{4}-\frac{\mathit{K1}}{4}+\frac{\mathit{K2}}{4}+\mathit{E_B}  & 0 & 0 & \frac{\mathit{K2}}{4} & 0 & -\frac{\mathit{K2}}{4} & 0 & 0 
\\
 0 & 0 & 0 & \frac{\mathit{J1}}{2}+\frac{\mathit{K1}}{4} & 0 & \frac{\mathit{J1}}{2}+\frac{\mathit{K1}}{4} & 0 & 0 & 0 & -\frac{\mathit{J1}}{4}-\frac{\mathit{J2}}{4}+\frac{\mathit{K1}}{4}+\frac{\mathit{K2}}{4} & \frac{\mathit{J2}}{2}+\frac{\mathit{K2}}{4} & 0 & \frac{\mathit{J2}}{2}+\frac{\mathit{K2}}{4} & 0 & 0 & 0 
\\
 -\frac{\mathit{K1}}{4} & 0 & 0 & \frac{\mathit{J1}}{2} & 0 & 0 & \frac{\mathit{J1}}{2}+\frac{\mathit{K1}}{4} & 0 & 0 & \frac{\mathit{J2}}{2}+\frac{\mathit{K2}}{4} & -\frac{\mathit{J1}}{4}-\frac{\mathit{J2}}{4}-\frac{\mathit{K1}}{4}-\frac{\mathit{K2}}{4} & 0 & \frac{\mathit{J2}}{2} & 0 & 0 & -\frac{\mathit{K2}}{4} 
\\
 0 & -\frac{\mathit{K1}}{4} & 0 & 0 & 0 & 0 & 0 & \frac{\mathit{J1}}{2}+\frac{\mathit{K1}}{4} & \frac{\mathit{K2}}{4} & 0 & 0 & \frac{\mathit{J1}}{4}-\frac{\mathit{J2}}{4}+\frac{\mathit{K1}}{4}-\frac{\mathit{K2}}{4}-\mathit{E_B}  & 0 & \frac{\mathit{J2}}{2} & \frac{\mathit{J2}}{2}+\frac{\mathit{K2}}{4} & 0 
\\
 \frac{\mathit{K1}}{4} & 0 & 0 & 0 & 0 & \frac{\mathit{J1}}{2} & \frac{\mathit{J1}}{2}+\frac{\mathit{K1}}{4} & 0 & 0 & \frac{\mathit{J2}}{2}+\frac{\mathit{K2}}{4} & \frac{\mathit{J2}}{2} & 0 & -\frac{\mathit{J1}}{4}-\frac{\mathit{J2}}{4}-\frac{\mathit{K1}}{4}-\frac{\mathit{K2}}{4} & 0 & 0 & \frac{\mathit{K2}}{4} 
\\
 0 & \frac{\mathit{K1}}{4} & 0 & 0 & 0 & 0 & 0 & \frac{\mathit{J1}}{2}+\frac{\mathit{K1}}{4} & -\frac{\mathit{K2}}{4} & 0 & 0 & \frac{\mathit{J2}}{2} & 0 & \frac{\mathit{J1}}{4}-\frac{\mathit{J2}}{4}+\frac{\mathit{K1}}{4}-\frac{\mathit{K2}}{4}-\mathit{E_B}  & \frac{\mathit{J2}}{2}+\frac{\mathit{K2}}{4} & 0 
\\
 0 & 0 & \frac{\mathit{K1}}{4} & 0 & -\frac{\mathit{K1}}{4} & 0 & 0 & \frac{\mathit{J1}}{2} & 0 & 0 & 0 & \frac{\mathit{J2}}{2}+\frac{\mathit{K2}}{4} & 0 & \frac{\mathit{J2}}{2}+\frac{\mathit{K2}}{4} & \frac{\mathit{J1}}{4}-\frac{\mathit{J2}}{4}-\frac{\mathit{K1}}{4}+\frac{\mathit{K2}}{4}-\mathit{E_B}  & 0 
\\
 0 & 0 & 0 & \frac{\mathit{K1}}{4} & 0 & -\frac{\mathit{K1}}{4} & 0 & 0 & 0 & 0 & -\frac{\mathit{K2}}{4} & 0 & \frac{\mathit{K2}}{4} & 0 & 0 & \frac{3 \mathit{J1}}{4}+\frac{3 \mathit{J2}}{4}+\frac{\mathit{K1}}{4}+\frac{\mathit{K2}}{4}-2 \mathit{E_B}  
\end{array}
\end{pmatrix}
$}
\label{eq:tetrahedron}
\end{equation}
\end{widetext}

\FloatBarrier

\bibliography{KitaevCluster}

\end{document}